\documentclass[leqno,a4paper,12pt]{article}
\usepackage[centertags]{amsmath}
\usepackage{amsfonts}
\usepackage{amssymb}
\usepackage{multirow}
\usepackage{amsthm}
\usepackage{nicefrac}
\usepackage{pifont}
\usepackage[ansinew]{inputenc}
\usepackage{graphicx}

\DeclareMathOperator{\supp}{supp}
\DeclareMathOperator{\tri}{tri}

\begin{document}

\title{Incidence, recovery and prevalence of infectious diseases: 
non-parametric disease model and application to influenza in Germany}
\author{Ralph Brinks\\German Diabetes Center}
\maketitle

\begin{abstract}
In this work we describe a non-parametric disease model that links the temporal change of the
prevalence of an infectious disease to the incidence and the recovery rates. 
The model is only based on the common epidemiological measures incidence and recovery rate. 
As an application, the model is used to calculate the prevalence of influenza in Germany for
a hypothetical birth cohort during 2001 and 2013.
\end{abstract}

\section*{Introduction} 
In mathematically modelling infectious diseases, often compartment models are used. 
Compartment models divide the population under consideration into disjunct sets 
of individuals with the same biological characteristics. Prominent examples 
in infectious disease modelling are the \emph{SI}, \emph{SIS} and \emph{SIRS} models,
see for example \cite{Ear08, Kee08}. The
models have in common that they depend on one or more parameters. 
For instance, all these models need a parameter, mostly called \emph{transmission rate} 
$\beta,$ that describes how effective contacts between susceptible and infected persons are 
with respect to spreading the disease. Biological, chemical and physical properties of 
infectious agents as well 
as the behaviour of hosts, susceptible or infected, lead to a variety of possible values 
of $\beta.$ Even within the same class of disease the transmission characteristics may
vary considerably, which was shown for example in influenza \cite{teB13}. This may impose
practical problems in estimating and predicting the parameters.

\bigskip

In this work we analyse the temporal dynamics of the prevalence of infectious diseases
in a non-parametric way. The temporal change of the prevalence is expressed in terms
of the incidence and the recovery rate.

\section*{SD-Model} \label{sec:SD} 
\begin{figure}[ht]
  \centering
  \includegraphics[keepaspectratio,width=0.9\textwidth]{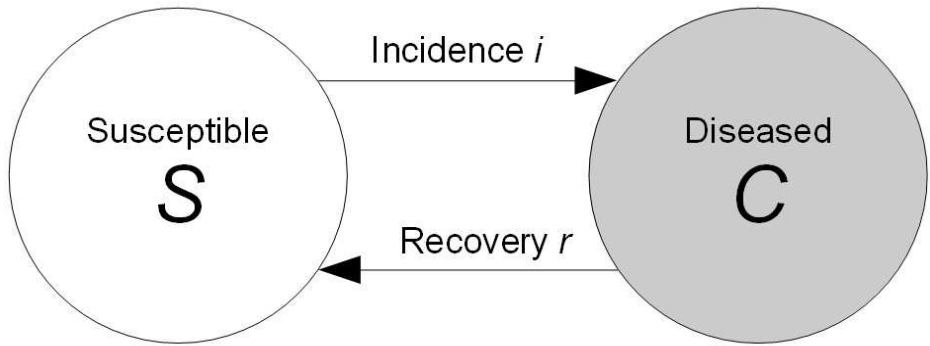}
\caption{Disease model with two states and the corresponding transition rates. 
Persons in the state \emph{Susceptible} are healthy with respect to the disease under 
consideration. After onset of the disease they change into the \emph{Diseased} state. 
Later they recover and return to the \emph{Susceptible} state.}
\label{fig:CompModel}
\end{figure}

We start with a simple compartment model that divides the population into those who
are not infected (suceptible), and those who are diseased (Figure 
\ref{fig:CompModel}). The numbers of persons in the states \emph{Susceptible} and \emph{Diseased} 
are denoted by $S$ and $C$ (cases). The transition rates between the 
states are the incidence rate $i$ and the recovery rate $r$, which depend on
the time variable $t.$

The equations characterising the changes of $S$ and $C$ in the compartment model 
of Figure \ref{fig:CompModel} are:
\begin{subequations}\label{e:balance}
\begin{align}
\frac{\mathrm{d} S}{\mathrm{d} t} &= - i \, S + r \, C\\  
\frac{\mathrm{d} C}{\mathrm{d} t} &=   i \, S - r \, C.   
\end{align}
\end{subequations}

By applying the quotient rule to the prevalence $p = \tfrac{C}{S+C}$ 
and inserting these equations we get the following scalar ordinary differential equation (ODE)

\begin{equation}\label{e:dp}
\frac{\mathrm{d} p}{\mathrm{d} t} = (1-p) \, i - p \, r.
\end{equation}

The linear ODE \eqref{e:dp} shows that the temporal change $\tfrac{\mathrm{d} p}{\mathrm{d} t}$
of the prevalence is a convex combination of the incidence rate $i$ and the recovery rate
$r$. The solution of \eqref{e:dp} with the initial condition $p(t_0) = p_0$ is

\begin{equation}\label{e:solution}
p(t) = \exp \bigl ( -G(t) \bigr) \left \{p_0 + \int \limits_{t_0}^t 
i(\tau) \exp \bigl ( G(\tau) \bigr ) \mathrm{d} \tau \right \},
\end{equation}

where $$G(t) = \int \limits_{t_0}^t i(\tau) + r(\tau) \, \mathrm{d}\tau.$$

\paragraph{Remark 1:} If $i$ and $r$ are constant and the disease is in equilibrium, 
i.e. $\tfrac{\mathrm{d} p}{\mathrm{d} t} = 0,$ Equation \eqref{e:dp} in case of $p \neq 1$
reads as
\begin{equation}\label{e:equi}
\frac{p}{1-p} = \frac{i}{r}.
\end{equation}
This is the well-known result that the prevalence odds $\tfrac{p}{1-p}$ equals the product of 
incidence and mean duration of the disease.

\paragraph{Remark 2:}
For later use we define the triangle function $\tri_{a, b, h}.$ Let $a < b$  and $h >0,$ then
set
\[\tri_{a, b, h} (t) := 
\begin{cases}
   h \cdot \left ( 1 - \frac{2 \cdot \vert t - \frac{a+b}{2} \vert} {b-a} \right ) & \textnormal{for } a < t < b\\
   0 & \textnormal{else.}
\end{cases}
\]
The function has a triangular shape with a peak of height $h$ at $t = \tfrac{a+b}{2}.$ An example
of a triangle function is the red curve in Figure \ref{fig:Triangle}.

\section*{Examples} 

In this section we illustrate Equations \eqref{e:dp} and \eqref{e:solution} by some examples.

\subsection*{Example 1}
\begin{figure}[ht]
  \centering
  \includegraphics[keepaspectratio,width=0.95\textwidth]{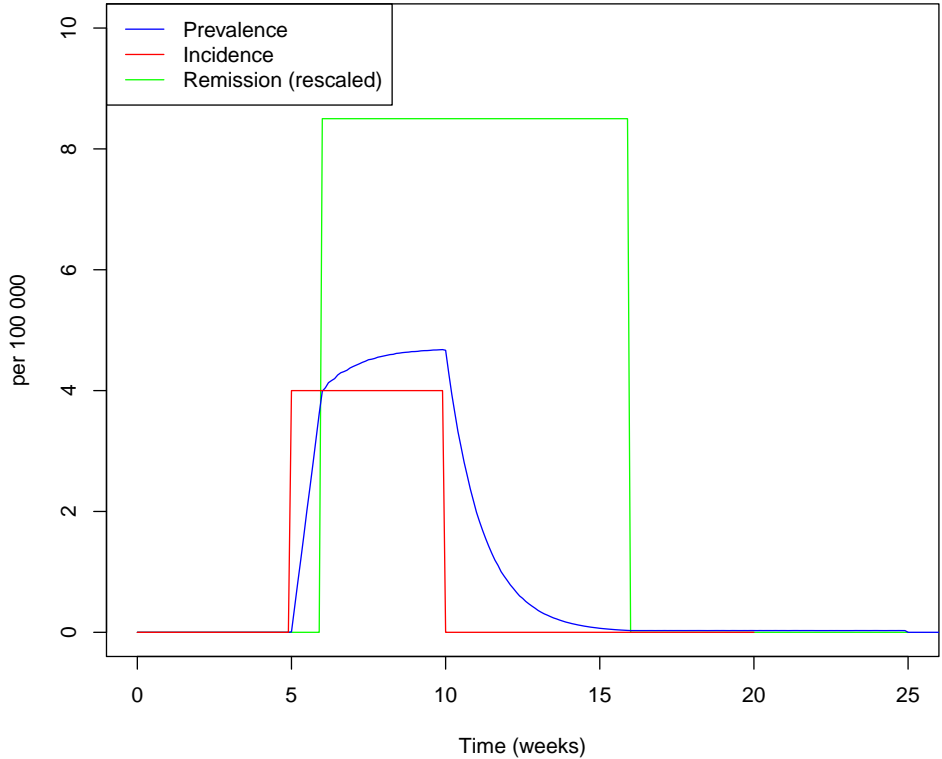}
\caption{Time course of the incidence (red) and recovery rate (green, rescaled by multiplying
with $10^{-4}$) in Example 1. The resulting prevalence according to Equation \eqref{e:solution} is
is the blue curve.}
\label{fig:Rectangle}
\end{figure}

The first example assumes a rectangular time course of the incidence (Figure \ref{fig:Rectangle}).
The \emph{support} of the incidence (i.e., the set $\supp(i) := \{t \,| ~i(t) > 0\}$) is 
$(5, 10)$, the support of the recovery rate is $\supp(r) = (7, 17)$. On these intervals 
the values of the incidence and recovery are assumed to be $4 \cdot 10^{-5}$ and 0.85 
(per week), respectively.
The resulting prevalence (calculated by numerically integrating Equation \eqref{e:solution}) is
shown as blue curve in Figure \ref{fig:Rectangle}.

\subsection*{Example 2}
\begin{figure}[ht]
  \centering
  \includegraphics[keepaspectratio,width=0.8\textwidth]{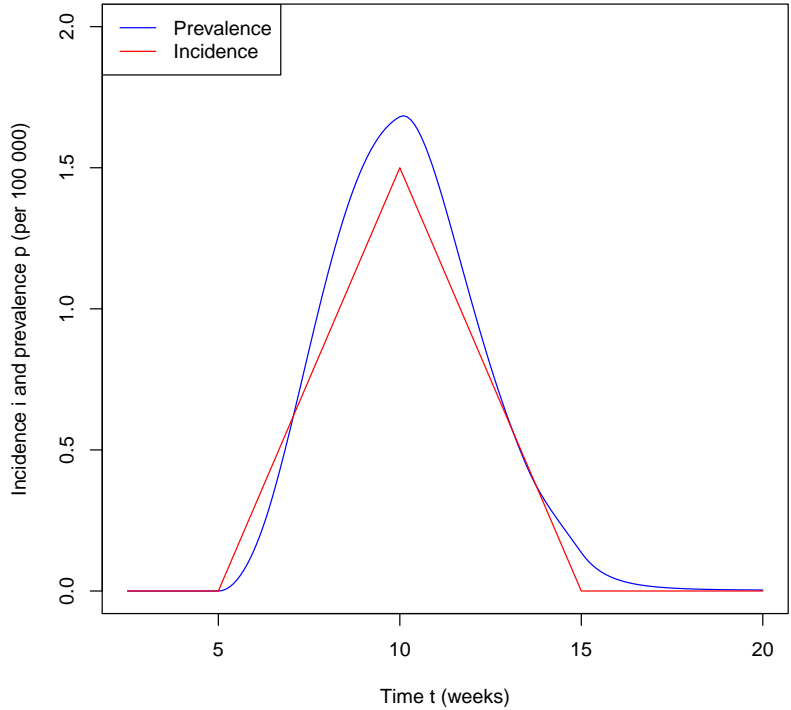}
\caption{Time course of the incidence rate (red) and prevalence (blue) in Example 2. 
Note the delay between the incidence and the prevalence.}
\label{fig:Triangle}
\end{figure}

As we shall see in the next section, the time course of the incidence in a wave of influenza does 
not have a rectangular shape. It is (approximately) symmetric and has a peak in the middle. 
Compared to the previous example, a triangular shape of the incidence is more realistic. 
We assume a wave of influenza having the incidence as shown by the red curve of Figure 
\ref{fig:Triangle}: $i = \tri_{5, 15, h}$ with $h = 1.5 \cdot 10^{-5}$ (per week). 
The recovery rate $r$ is assumed to be a triangle function, too, with 
$\supp(r) = (6.5, 20)$ and peak height 1.7 (per week). The time course of the associated 
prevalence (blue line in Figure \ref{fig:Triangle}) has been calculated by Equation \eqref{e:solution}.

From Figure \ref{fig:Triangle} it is apparent that the prevalence starts to increase later than
the incidence. At about week 7, the prevalence has overtaken the incidence. In week 10 the incidence
peaks at $1.5 \cdot 10^{-5}$ (per week), whereas the prevalence peaks at about 0.3 weeks later at the 
value $1.68 \cdot 10^{-5}.$ In summary, we can see that the prevalence is delayed compared to the 
incidence and overshoots the peak of the incidence.

\subsection*{Example 3: Equilibrium}
To illustrate Equation \eqref{e:equi}, we have chosen $i(t) = 10^{-5}$ for $t \ge 5$
and $r(t) = 0.2$ for $t \ge 10$. The associated duration of the disease is $\tfrac{1}{0.2} = 5.$
\begin{figure}[ht]
  \centering
  \includegraphics[keepaspectratio,width=0.8\textwidth]{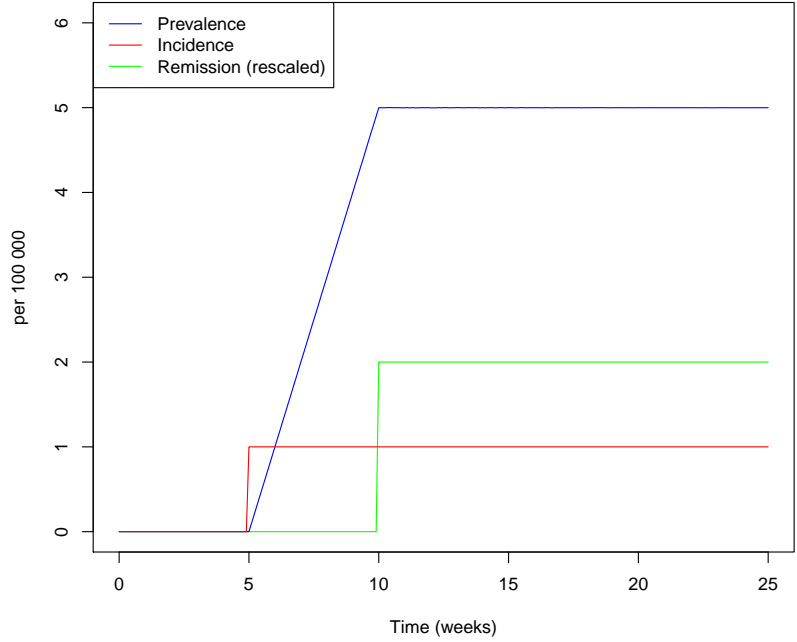}
\caption{Time course of the incidence rate (red), recovery rate (green, rescaled by multiplying
with $10^{-4}$) and prevalence (blue) in Example 3.}
\label{fig:Stationary}
\end{figure}
Beginning at $t = 10$ the prevalence is constant (equilibrium). 
It holds $\tfrac{p(t)}{1-p(t)} = 5 \cdot 10^{-5}$ for all
$t \ge 10$. Figure \ref{fig:Stationary} shows the course of the associated prevalence (blue). 

\subsection*{Example 4}
For use in the next section we solve the following problem: given the
triangular incidence $i = \tri_{a_1, b_1, h_1}, ~ h_1 > 0, a_1 < b_1,$ what has to be the minimal 
$h_2$ in a triangular recovery $r = \tri_{a_2, b_2, h_2}, ~ h_2 > 0, a_2 < b_2$ and $a_1 < a_2, b_1 < b_2,$ 
such that $p(T) = 0$ 
for all $T \ge b_2$? With other words: what is the minimal peak height $h_2 > 0$ of a triangular 
recovery rate $r$ that follows after a triangular incidence $i$ 
with height $h_1$ such that the disease is eradicated at 
$T \ge b_2.$

\bigskip

For $T \ge b_2$ it holds 
$$p(T) = \exp \bigl ( -G(T) \bigr) \int \limits_{a_1}^{b_1} 
\tri_{a_1, b_1, h_1}(\tau) \exp \bigl ( G(\tau) \bigr ) \mathrm{d} \tau$$
with
$$G(t) = \int \limits_{a_1}^t \tri_{a_1, b_1, h_1}(\tau) + \tri_{a_2, \, b_2, \, h_2}(\tau) \, \mathrm{d}\tau.$$

It is easy to see that $p(T) = p(b_2)$ for all $T \ge b_2$. Thus, we may speak of the 
\emph{terminal prevalence}. As the terminal prevalence $p(T)$ is the product of two 
positive factors, the prevalence is positive 
for all $T \ge b_2.$ Hence, the only aim we can achieve is to bring $p(T)$ below a prescribed
threshold. That this is possible, can be seen by the following calculation
\begin{align*}
p(T) &=   \exp(-G(T)) \int_{a_1}^{b_1} \tri_{a_1, b_1, h_1}(\tau) \exp \bigl ( G(\tau) \bigr ) \mathrm{d} \tau\\
     &\le \exp \bigl( G(b_1) - G(T) \bigr) \int_{a_1}^{b_1} \tri_{a_1, b_1, h_1}(\tau) \, \mathrm{d} \tau \\
     &= \frac{1}{2} \, h_1 \, (b_1-a_1) \, \exp \bigl( G(b_1) - G(T) \bigr).
\end{align*}
The inequality holds true, because $G$ is monotonically increasing. From
$$G(b_1) - G(T) = - \int \limits_{\max(b_1, a_2)}^{b_2} \tri_{a_2, \, b_2, \, h_2}(\tau) \, \mathrm{d}\tau =
 - \frac{1}{2} \, h_2 \, \bigl( b_2 - \max(b_1, a_2) \bigr)$$ it follows that
$p(T) \rightarrow 0$ as $h_2 \rightarrow \infty.$

\bigskip

For a given incidence $i = \tri_{a_1,b_1,h_1}$ and $a_2, b_2$ we are interested 
in the minimal $h$ such that the terminal prevalence $p(T)$ is below a prescribed threshold. 

\bigskip

\begin{figure}[ht]
  \centering
  \includegraphics[keepaspectratio,width=0.8\textwidth]{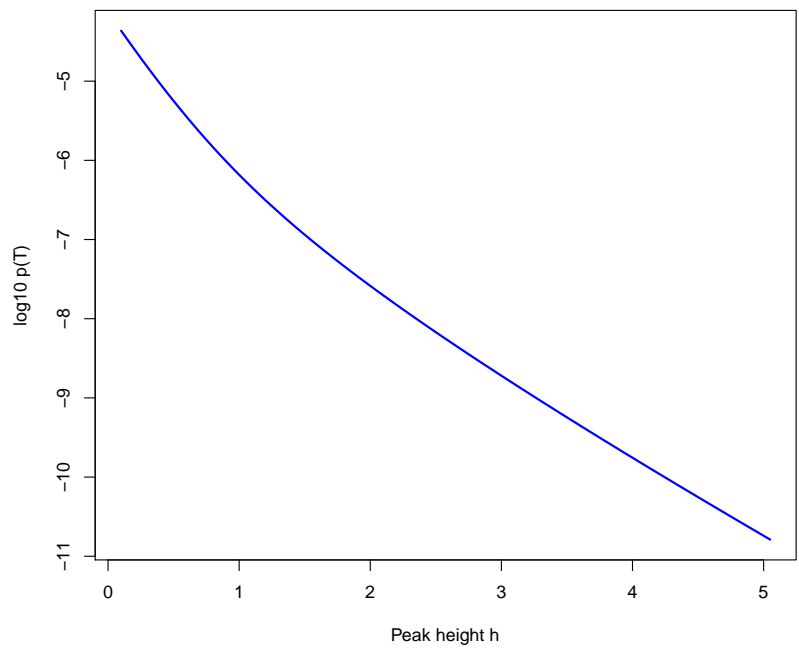}
\caption{Logarithm of the terminal prevalence $p(T), ~T \ge b_2,$ as a function of the
peak height $h$.}
\label{fig:TerminalP}
\end{figure}

To solve this problem, we examine the function 
$$\mathcal{H}_{a_1, b_2, h_1, a_2, b_2}: h \mapsto p(T) = \int \limits_{a_1}^{b_1} 
\tri_{a_1, b_1, h_1}(\tau) \exp \bigl ( G(\tau) - G(T) \bigr ) \mathrm{d} \tau,$$ where 
$G(t) = \int_{a_1}^{t} \tri_{a_1,\, b_1,\, h_1} (\tau) + \tri_{a_2, \, b_2, \, h} (\tau) \, \mathrm{d} \tau.$
Figure \ref{fig:TerminalP} gives an example of the terminal prevalence $p(T)$
for $a_1=5, ~b_1=15, ~h_1=1.5 \cdot 10^{-5}$ (incidence in Figure \ref{fig:Triangle}) and $a_2 = 6.5, ~
b_2 = 20.$

\bigskip

We will choose the threshold $h$ such that the terminal prevalence is one per mille of
the peak incidence, i.e. $p(T) \le \tfrac{h_1}{1000}.$ Then, we assume that the wave of 
influenza is eradicated after $b_2.$ The corresponding peak height $h_2$ in case of 
$a_1=5, ~b_1=15, ~h_1=1.5 \cdot 10^{-5}$ and $a_2=6.5, ~b_2=20$ is $h = 2.20.$

\clearpage

\section*{Influenza in Germany 2001-13} 
\begin{figure}[ht]
  \centering
  \includegraphics[keepaspectratio,width=0.95\textwidth]{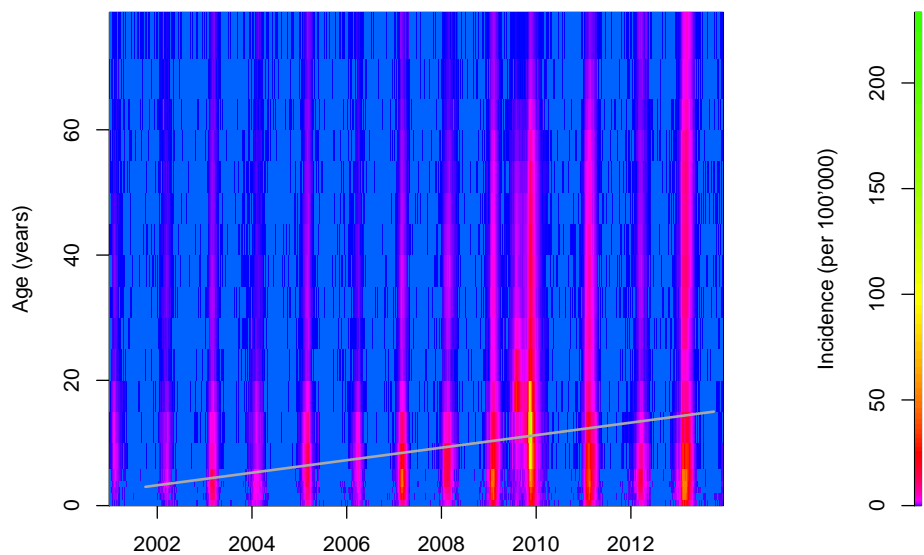}
\caption{Age-specific incidence of influenza (per 100 000) in Germany 2001-2013. The
whitish line in the lower part represents the trajectory of a hypothetical cohort.}
\label{fig:Influenza}
\end{figure}

Figure \ref{fig:Influenza} shows the incidence of influenza in Germany from 2001 to
2013. The abscissa and ordinate represent calendar time and age, respectively. The
colour indicates the incidence rate, the associated numerical values are coded as shown
in the rightmost part of Figure \ref{fig:Influenza}. By incidence we mean the
incidence reported to the national influenza register at the Robert-Koch-Institute, 
\cite{Sur13}. In Germany, all confirmed influenza cases statutorily have to be 
reported to the Robert-Koch-Institute (influenza A, B, C according to the reference 
definition).

From Figure \ref{fig:Influenza} it becomes apparent that influenza usually 
appears in the first quarter of the year and vanishes (nearly) completely from the second 
to the fourth quarter. An exception is the epidemic in the last quarter of 2009. Then,
the swine flu (H1N1 influenza) became pandemic. We also see that not all age groups are
affected equally from one wave of influenza to the other. It seems that as calendar time
progresses, the more older age groups get involved.

\begin{figure}[ht]
  \centering
  \includegraphics[keepaspectratio,width=0.95\textwidth]{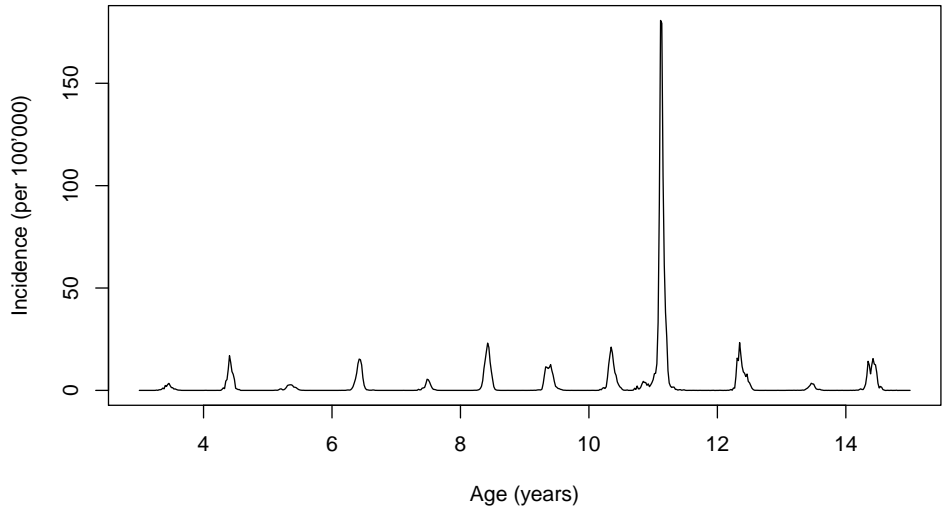}
\caption{Incidence of influenza for a birth cohort followed up along the whitish line
in Figure \ref{fig:Influenza}.}
\label{fig:Cohort}
\end{figure}

In Figure \ref{fig:Influenza} a whitish line is visible in the lower third of the image.
This is the trajectory of a hypothetical birth cohort born in September 1998, which has
been followed from September 2001 to September 2013. The values of the incidence rate along
the line is shown in Figure \ref{fig:Cohort}. The seasonal variability and the
enormous peak during the swine flu pandemic are clearly visible.

\begin{figure}[ht]
  \centering
  \includegraphics[keepaspectratio,width=0.95\textwidth]{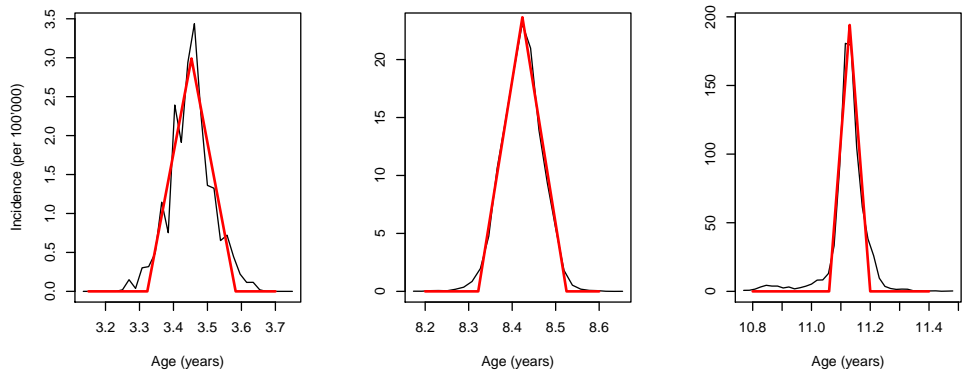}
\caption{Three influenza waves of the birth cohort in Figure \ref{fig:Cohort}: spring 2002 (left),
spring 2007 (middle) and autumn 2009 (right). The black curves are the raw incidence rates as reported
to the Robert-Koch-Institute \cite{Sur13}. The triangle functions (red) are the approximated incidence 
rates. Note the different scalings of the ordinate.}
\label{fig:IncidenceThreeWaves}
\end{figure}

During follow-up, the birth cohort faces twelve waves of influenza with different
intensities (see Figure \ref{fig:Cohort}). Three of them are analysed in more detail: 
the wave with relatively low incidence at 3.5 years of age (spring 2002), the
moderate wave at age 8.5 (spring 2007) and the swine flu at age 11.3 (autumn 2009). The
corresponding incidences are shown as black curves in Figure \ref{fig:IncidenceThreeWaves}.

The raw incidence data are approximated by triangle functions $i_k = \tri_{a_k, b_k, y_k},$ $k=1,2,3.$ 
These have been calculated by an ordinary least squares approach. As in Example 4 (see above) 
we assume triangle functions for the recovery rates $r_k, ~k=1, 2, 3.$ The support of
the recovery rate $r$ is assumed to be $a_2 = a_1 + 3/365.25$ and $b_2 = b_1 + 17/365.25.$ 
This corresponds to a mean delay of 10 with range 3-17 (days). The peak heights $h_k$ of the 
recovery rates $r_k$ are calculated as in Example 4 by forcing the terminal prevalence to 
be less than one per mille of the peak incidence. The results are presented in Table \ref{tab:waves}.

\begin{table}[ht!]
\caption{Analysed influenza waves of the birth cohort and characteristics of the
triangular incidence and recovery rates.}
\label{tab:waves}
\begin{center}
\begin{tabular}{c | c  c | c  c  }
\hline\noalign{\smallskip}
Influenza   & \multicolumn{2}{|c|}{Incidence $i$}       & \multicolumn{2}{c}{Recovery $r$}  \\
wave        & Support $\supp(i)$ &   Peak   height      & Support $\supp(r)$ & Peak  height  \\ \hline
Spring 2002 & $(3.323, 3.583)$   & $2.99 \cdot 10^{-5}$ & $(3.350, 3.610)$   &   190.35      \\ 
Spring 2007 & $(8.322, 8.525)$   & $23.7 \cdot 10^{-5}$ & $(8.349, 8.552)$   &   167.61      \\ 
Autumn 2009 & $(11.060, 11.200)$ & $ 194 \cdot 10^{-5}$ & $(11.087, 11.227)$ &   137.85      \\ 
\noalign{\smallskip}\hline
\end{tabular}
\end{center}
\end{table}

\clearpage

\begin{figure}[ht]
  \centering
  \includegraphics[keepaspectratio,width=0.95\textwidth]{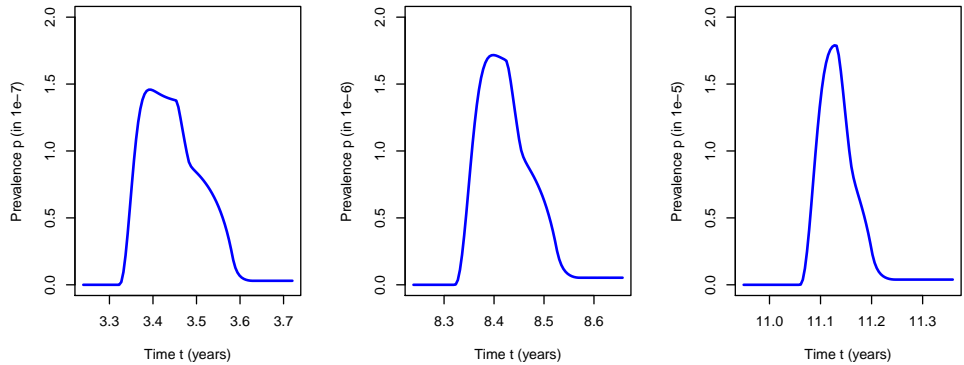}
\caption{Prevalence of influenza in the birth cohort during the three analysed waves: spring 2002 (left),
spring 2007 (middle) and autumn 2009 (right). 
Note the different scalings of the ordinate.}
\label{fig:PrevalenceThreeWaves}
\end{figure}

The time courses of the prevalence during the three waves of influenza are depicted in Figure
\ref{fig:PrevalenceThreeWaves}. The shapes are very similar but the peak values differ considerably.

During the waves in spring 2002 and 2007 the peak prevalence was $1.458 \cdot 10^{-7}$ and $1.716 
\cdot 10^{-6},$ respectively. The maximum of the prevalence during the swine flu epidemic in autumn
was $1.790 \cdot 10^{-5}.$ Roughly speaking, the three peak prevalences differ
by about one magnitude. This is consistent with the incidence rates, which approximately differ by a
factor of 10.

\clearpage

\section*{Discussion} 
In this work we have described a disease model that links the temporal change of the
prevalence to the incidence and the recovery rates. The model is non-parametric in the
sense that it does not depend on biological, behavioural or disease-specific parameters. 
It is only based on the common epidemiological measures incidence and recovery rate. In that respect 
the model is very flexible and is not restricted to a specific class of infectious disease.

\bigskip

After the introduction of the disease model, the characterising equations and some examples, 
the model has been applied to
incidence of influenza in Germany during 2001-2013. The incidence data stem 
from the Robert-Koch-Institute, which is the official authority each confirmed case of influenza 
in Germany by law has to be reported to. Since data about recovery rates are not published, assumptions had to be
made. With these assumptions the prevalence of influenza during three waves has been calculated for
a hypothetical birth cohort. During the three waves of influenza, the resulting prevalence portions
in the birth cohort are low. There are mainly two reasons: the first is the short duration of symptoms of averagely 10 days 
(range 3-17 days), which implies a high recovery rate with onset soon after the start of the wave
of influenza. The second reason lies in the data itself. Presumably, the cases reported 
to the Robert-Koch-Institute are only the most severe cases. It is very likely that a lot of patients
with the symptoms of influenza have not been examined by a medical doctor at all, or 
have not been examined in detail (for example by PCR). Those cases have not been \emph{confirmed influenza
cases} that statutorily have to be reported. Thus, they are not covered by the incidence rates in this article. 
The fraction of unreported cases is difficult to access and is beyond the scope of this work. 
Although the equations are mathematically correct in the context they were developed for (no mortality,
no migration), due to this coverage issue the calculated prevalence portions 
have to be interpreted very carefully.

\bigskip

Another limitation of this article lies in the mathematical models for the incidence and
recovery rates. Here, we have used rectangle and triangle functions. In countries with seasonal
waves of influenza, incidence and recovery rates vanish in certain periods. Thus, 
in modelling single waves of influenza, functions with
bounded support would be preferable. Here, we have chosen triangle and rectangle functions, but
other functions may be possible as well, for example B-splines. In regions where influenza is present
during the whole year, incidence and remission do not have bounded support.

\bigskip

So far, the model in Figure \ref{fig:CompModel} does not include the impact of mortality. 
One may do so by changing the equations in \eqref{e:balance}:
\begin{eqnarray*}
\frac{\mathrm{d} S}{\mathrm{d} t} &=& - i \, S + r \, C - m_0 \, S\\ 
\frac{\mathrm{d} C}{\mathrm{d} t} &=&   i \, S - r \, C - m_1 \, C. 
\end{eqnarray*}
Then, Equation \eqref{e:dp} changes to
\begin{equation}\label{e:dp2}
\frac{\mathrm{d} p}{\mathrm{d} t} = (1-p) \, i - p \, r - p \, (1-p) \, \Delta m,
\end{equation}
with $\Delta m = m_1 - m_0$, \cite{Bri13}. Note that Equation \eqref{e:dp2} becomes Equation \eqref{e:dp}
in case of $m_1 = m_0$. Here an interesting point becomes obvious: Equation \eqref{e:dp} 
is a consequence of Equation \eqref{e:balance}. However, while
Equation \eqref{e:balance} implyies $\frac{\mathrm{d} (S + C)}{\mathrm{d} t} = 0,$
i.e. the population size remains constant, Equation \eqref{e:dp} does not imply this. 
As it is a special case of Equation \eqref{e:dp2}, Equation \eqref{e:dp} holds true 
in the presence of mortality with $m_1 = m_0.$

{}

\bigskip

\emph{Contact:} \\
Ralph Brinks \\
German Diabetes Center \\
Auf'm Hennekamp 65 \\
D- 40225 Duesseldorf\\
\verb"ralph.brinks@ddz.uni-duesseldorf.de"
\end{document}